\documentclass[aps,prl,twocolumn,showpacs,
groupedaddress]{revtex4}  
\usepackage{graphicx}  
\usepackage{dcolumn}   
\usepackage{bm}        
\usepackage{amssymb}   
\usepackage[]{units}
\usepackage{amsmath}

\begin{document}



\title{Relativistic Superfluids in Curved Spacetime}
                             
\author{Kristian Hauser A. Villegas}
 \email{khvillegas@gmail.com}
\date{\today}

\begin{abstract}
Superfluids under an intense gravitational field are typically found in compact stellar cores. Most treatments of these superfluids, however, are done using a flat spacetime background. In this paper, the effect of spacetime curvature on relativistic superfluids is investigated. The scalar-field superfluid in a background metric of a typical star core is considered first. It is found that the superfluid formed inside a compact object can not be of spherical shape. Explicit numerical calculation of the gravitational correction to the superfluid order parameter is performed for two specific examples with different boundary conditions. It is found that even in the weak-gravity limit, gravity can have a significant effect on superfluidity. The relativistic superfluids formed by  various fermion pairings are also considered. Two possible cases are considered in the mean-field treatment: antifermion-fermion and fermion-fermion pairings. The effective actions and the self-consistent equations are derived for both cases. An analytical expression for the first-order gravitational correction to the superfluid order parameter is given for the antifermion-fermion pairing; while, the analytical expressions for the matrix elements of the heat kernel operator, which is useful in curved-spacetime QFT calculations, is derived for the fermion-fermion pairing.

\end{abstract}

\pacs{}
\maketitle

\section{\label{sec:level1}I. Introduction}
Superfluids are ubiquitous inside compact stellar objects like neutron stars and the still-hypothetical quark stars\cite{Itoh70, Alford06, Jaikumar06, Masuda13, Endo14, Weber14}. Examples of these include neutron superfluid, superconducting proton, pion and kaon condensates, and the two-color superconducting and color-flavor-locked phases in quark matter. Most treatments of these superfluids are either extrapolations from a non-relativistic many-body theory or relativistic, but flat-spacetime quantum field theory\cite{Pethick99, Fukushima11, Alford08, Bailin84, Pisarski99}. At the core of these compact stellar objects, however, it is not possible to neglect the effect of gravity\cite{Teukolsky83}. Although General Relativity is accounted in the stellar structure calculations via the Tolman-Oppenheimer-Volkoff equations for hydrostatic equilibrium, the dense-matter properties, like superfluidity and various quark pairings, are usually calculated using a flat spacetime background. A consistent treatment should include the effects of General Relativity and treat the possible formation of superfluids, including color-superconductors and color-flavor locked phases in quark matter, in curved spacetime. The only study about the effects of gravity on a quark pairing that the author is aware of, treats the effect of the expanding universe on the color superconductor instead of the gravity produced by the dense quark matter itself\cite{Ebert07}.

This work fills this gap in literature and accounts the effect of spacetime curvature in the formation of various relativistic superfluids. Our aim is to lay down and derived the basic equations necessary for the treatment of superfluidity in curved spacetimes and to show that gravity has significant and non-trivial effects on the formation of superfluids. The analytical forms of effective actions, self-consistent equations, and the heat kernel operator that are relevant in the treatment of relativistic superfluids in curved spacetime background will be derived. Numerical calculations are provided for the scalar-field superfluid.

This paper is organized as follows. In Section II, the effect of gravity on the superfluidity of a relativistic real scalar field will be considered. This case is special as it is simple enough to enable us to do explicit numerical calculations. The first-order gravitational correction to the superfluid order parameter is calculated for two sample cases, differing in their boundary conditions. The following results are shown: i.) the superfluid inside a compact star can not be spherical, ii.) depending on the boundary conditions, gravity can enhance or destroy superfluidity and iii.) even in the weak-gravity regime, the gravitational correction to the superfluid order parameter can be as significant as $\sim 0.8\lambda$ where $\lambda$ is the strength of the gravitational perturbation

In Section III, the effective four-fermion interaction in curved spacetime will be derived by integrating out the scalar field in a boson-mediated fermion-fermion Yukawa interaction. The extension to the more realistic case of gauge vector boson-mediated fermion interactions will also be discussed. 

The two possible types of mean-field ansatz is considered in Section IV: the antifermion-fermion pairing and the fermion-fermion pairing. The effective actions that are quadratic in fermion field and the self-consistent equations will be derived for both types of pairings. The analytical expression for the first-order gravitational correction is derived for the antifermion-fermion case. The fermion-fermion pairing is further developed since this is the type of pairing that occurs in the color-superconducting state and the color-flavor locked phase in quark matter. Specifically, the explicit analytical forms of the Euclidean effective action and the matrix elements of the heat kernel operator, which are relevant in curved-spacetime quantum field theory calculations, will be calculated. 

Finally, concluding remarks will be given in Section V.

\section{II. Bosonic Superfluid}
Let us first consider the superfluidity of a massless real scalar field. The Wick-rotated euclidean action is taken to be 
\begin{eqnarray}
S_E[\phi,g_{\mu\nu},J]=\frac{1}{2}\int d^4x\sqrt{g}\big(-\hat{\psi}\square\hat{\psi}-\mu\hat{\psi}^2-\lambda\hat{\psi}^4\big)
\end{eqnarray}
where $\square\phi\equiv g^{-1/2}\partial_{\nu}(g^{1/2}g^{\mu\nu}\partial_{\mu}\hat{\psi})$ and $\hat{\psi}$ is the scalar field operator. 

We now introduce the superfluid order parameter
\begin{eqnarray}
\Psi(x)\equiv\langle\hat{\psi}(x)\rangle\neq 0
\end{eqnarray}
and the fluctuations about this mean-field value
\begin{equation}
\hat{\phi}(x)\equiv \hat{\psi}(x)-\Psi(x)
\end{equation}
which is assumed to be small.

We now expand the action (1) up to terms quadratic in the fluctuation $\hat{\phi}(x)$. The relevant terms are
\begin{eqnarray}
S^0_E&=&\frac{1}{2}\int d^4x\sqrt{g}\bigg[\Psi(-\square-\mu)\Psi-\lambda\Psi^4\bigg]\\
S^1_E&=&\frac{1}{2}\int d^4x\sqrt{g}\bigg[\hat{\phi}(-\square-\mu)\Psi-2\lambda\hat{\phi}\Psi^3\nonumber\\
&{}&{}+\Psi(-\square-\mu)\hat{\phi}-2\lambda\Psi^3\hat{\phi}\bigg]\\
S^2_E&=&\frac{1}{2}\int d^4x\sqrt{g}\bigg[\hat{\phi}(-\square-\mu-6\lambda\Psi^2)\hat{\phi}\bigg].
\end{eqnarray}
Equation (4) contributes a constant factor to the path integral and has no contribution in the calculation of expectation values. Equation (5) can be made to vanish by imposing that the order parameter $\Psi$ satisfies the Hartree self-consistent equation\cite{Fetter95}
\begin{equation}
(-\square-\mu-2\lambda\Psi^2)\Psi=0.
\end{equation}

Equation (7) is obtained by treating $\Psi$ as a variational parameter. Later, we will investigate the effect of spacetime curvature on the superfluid order parameter via this self-consistent equation.

The path integral of this action is evaluated by expanding in terms of the eigenfunctions of the eigenvalue equation
\begin{eqnarray}
\hat{F}\phi_n(x)\equiv [-\square -\mu-6\lambda\Psi^2(x)]\phi_n(x)=\lambda_n\phi_n(x)
\end{eqnarray}
obeying the orthonormal condition 
\begin{eqnarray}
\int d^4x\sqrt{g}\phi_m(x)\phi_n(x)=\delta_{mn}.
\end{eqnarray}

That is, 
\begin{eqnarray}
\phi(x)&=&\sum_nc_n\phi_n(x)\\
J(x)&=&\sum_nb_n\phi_n(x). 
\end{eqnarray}

The path integral measure becomes $\mathcal{D}\phi =\prod_n\frac{dc_n}{\sqrt{2\pi}}$. The $c_n$ fields are then shifted $c_n\rightarrow c_n+\frac{b_n}{2\lambda_n}$ to complete the square and the path integral can then be evaluated
\begin{eqnarray}
e^{-\Gamma_E}\equiv \int\mathcal{D}\phi e^{-S_E} =\exp\bigg\{-\frac{1}{8}\sum_n\frac{b^2_n}{\lambda_n}\bigg\}\bigg[\prod_n\lambda_n\bigg]^{-1/2}
\end{eqnarray}
and we have the Euclidean effective action
\begin{eqnarray}
\Gamma_E &=&\frac{1}{8}\sum_n\frac{b^2_n}{\lambda_n}+\frac{1}{2}\ln\det\hat{F}.
\end{eqnarray}

This can be rewritten back in terms of the source $J$ by inverting the eigenfunction expansion Eq.(11)
\begin{eqnarray}
\Gamma_E&=&\frac{1}{8}\int d^4x\sqrt{g(x)}\int d^4y\sqrt{g(y)}J(x)G_J(x,y)J(y)\nonumber\\
{}&{}&+\frac{1}{2}\ln\det\hat{F}
\end{eqnarray}
where
\begin{equation}
G_J(x,y)\equiv\sum_n\frac{\phi_n(x)\phi_n(y)}{\lambda_n}.
\end{equation}

The expression above provides the analytical form of the Euclidean effective action for the relativistic scalar-field superfluid. It is useful for calculating various correlation functions. For example, the two-point correlation function is given by
\begin{eqnarray}
\langle\phi(x)\phi(y)\rangle 
&=&e^{\Gamma_E}\bigg(\frac{2}{\sqrt{g(x)}}\frac{\delta}{\delta J(x)}\bigg)\nonumber\\
{}&{}&\times\bigg(\frac{2}{\sqrt{g(y)}}\frac{\delta}{\delta J(y)}\bigg)e^{-\Gamma_E}\bigg |_{J=0}\\
&=&-G_J(x,y).
\end{eqnarray}

We also note that from Eq.(14) we have
\begin{eqnarray}
\langle\phi(x)\rangle =e^{\Gamma_E}\bigg(\frac{2}{\sqrt{g(x)}}\frac{\delta}{\delta J(x)}\bigg)e^{-\Gamma_E}\bigg |_{J=0}=0
\end{eqnarray}
which is consistent with our definition of order parameter and fluctuations (2) and (3).

We now investigate the effect of gravity on the superfluid order parameter by considering the Hartree self-consistent equation (7). 

In flat spacetime, there is a macroscopic occupation in the single-particle ground state as the fugacity $z\equiv\exp (\beta\mu)$ goes to unity, or in terms of the chemical potential, $\mu\rightarrow 0$. Since we are interested in the case where the condensate order parameter is non-zero, we therefore take $\mu=0$ in the following discussion.

The self-consistent equation (7) can be solved perturbatively if it is assumed that the gravity and the coupling $\lambda$ are sufficiently weak. There are two sources of perturbations: the perturbation from the flat metric $g_{\mu\nu}=\eta_{\mu\nu}+h_{\mu\nu}$ and from the scalar field coupling $\lambda$. It is assumed that these two have comparable contributions. We can then expand $\Psi(x)=\Psi^{(0)}(x)+\Psi^{(1)}(x)+\cdot\cdot\cdot$ and $\square = \square^{(0)}+\square^{(1)}+\cdot\cdot\cdot$ where the superscripts $(0)$, $(1)$ denote zeroth order and first order in $\lambda$ or $h_{\mu\nu}$, respectively. 

Explicitly, the Laplacian operators are
\begin{eqnarray}
\square^{(0)}\Psi &\equiv &\partial^{\mu}\partial_{\mu}\Psi\\
\square^{(1)}\Psi &\equiv &\partial_{\mu}(h^{\mu\nu}\partial_{\nu}\Psi-\frac{1}{2}\delta_{\rho\sigma}h^{\rho\sigma}\delta^{\mu\nu}\partial_{\nu}\Psi)\nonumber\\
{}&{}&+\frac{1}{4}\delta_{\rho\sigma}h^{\rho\sigma}\partial^{\mu}\partial_{\mu}\Psi.
\end{eqnarray}

It is convenient to recast Eqs.(19) and (20) in an orthonormal coordinate instead of the general coordinate. Doing this, and considering only a static superfluid, then in zeroth order, Eq.(7) is simply the Laplace equation familiar in Classical electromagnetic theory
\begin{equation}
\nabla^2\Psi^{(0)}=0.
\end{equation}

The solution inside some sphere whose radius we take to be unity is
\begin{eqnarray}
\Psi^{(0)}(r,\theta,\phi)=\sum_{l=o}^{\infty}\sum_{m=-l}^{l}A_{lm}r^lY_{lm}(\theta,\phi)
\end{eqnarray}
where $A_{lm}$ are constant coefficients to be determined via the boundary condition and $Y_{lm}(\theta,\phi)$ are the spherical harmonics.

If the boundary condition is specified on the sphere of radius unity
\begin{eqnarray}
\Psi^{(0)}(r=1,\theta,\phi)=f(\theta,\phi)
\end{eqnarray}
for some given function $f(\theta,\phi)$, then the coefficients are given by
\begin{eqnarray}
A_{lm}=\int_0^{2\pi}d\phi\int_0^{\pi}d\theta\sin\theta Y^*_{lm}(\theta,\phi)f(\theta,\phi).
\end{eqnarray}

Note that if the superfluid forms a ball concentric to the star, then there is a spherical surface, the boundary between the superfluid and the normal state, for which $f(\theta,\phi)=0$. Equations (22) and (24) then gives $\Psi^{(0)}=0$ which is inconsistent with our assumption that the order parameter is non-zero. We therefore conclude the following: the shape of the superfluid inside a star's core can \textit{not} be in the shape of a ball. That is, the geometry of the superfluid-normal boundary is not spherical but necessarily contains bumps or lobes. This conclusion is unchanged even if the higher order corrections are included.

The first order term in the perturbative expansion of (7) is
\begin{eqnarray}
\nabla^2\Psi^{(1)}&=&-\lambda^{-1}\square^{(1)}\Psi^{(0)}-2(\Psi^{(0)})^3\\
&\equiv &-\rho_g(\mathbf{r})-\rho_c(\mathbf{r}),
\end{eqnarray}
where in the last line, we have separated the purely gravitational contribution, $\rho_g(\mathbf{r})\equiv \lambda^{-1}\square^{(1)}\Psi^{(0)}$, and the self-consistent contribution, $\rho_c(\mathbf{r})\equiv 2(\Psi^{(0)})^3$. The latter arises from the fact that we are solving the self-consistent Hartree equation iteratively.

Since the Laplacian operator is linear, we can similarly separate the first-order correction to the order parameter into gravity and self-consistent parts
\begin{eqnarray}
\Psi^{(1)}=\Psi^{(1)}_g+\Psi^{(1)}_c
\end{eqnarray}
so that $\Psi^{(1)}_g$ is now the solution to the Poisson equation
\begin{eqnarray}
\nabla^2\Psi^{(1)}_g=-\rho_g(\mathbf{r}).
\end{eqnarray}

In terms of the Green's function, it is given by
\begin{eqnarray}
\Psi^{(1)}(\mathbf{r})&=&\frac{1}{4\pi}\int_{V}\rho_g(\mathbf{r'})G(\mathbf{r},\mathbf{r'})d^3\mathbf{r'}\nonumber\\
{}&{}&-\frac{1}{4\pi}\oint f(\theta',\phi')\frac{\partial G}{\partial r'}da'
\end{eqnarray}
where the second integral is evaluated on the surface where the boundary condition is specified.

The Green's function expansion in spherical coordinates is
\begin{eqnarray}
G(\mathbf{r},\mathbf{r'})&=&4\pi\sum_{l=0}^{\infty}\sum_{m=-l}^lY^*_{lm}(\theta',\phi')Y_{lm}(\theta,\phi)\nonumber\\
{}&{}&\times\frac{r^l_<}{2l+1}\bigg(\frac{1}{r^{l+1}_>}-r^l_>\bigg)
\end{eqnarray}
where $r_<(r_>)$ is the lesser(greater) of $r$ and $r'$.

Let us now consider explicit examples. Due to the appearance of Laplace and Poisson equations, we consider examples of boundary conditions that are already familiar in classical electromagnetic theory textbooks. Although the reason for these choice of boundary conditions are purely pedagogical, by investigating such explicit examples, it will be seen that gravity has significant and non-trivial effects on superfluidity.

The perturbing metric is taken to be $h_{\mu\nu}=\mbox{diag}(2\Phi(r),-2\Phi(r),0,0)$; while, the flat part is written as $\eta_{\mu\nu}=\mbox{diag}(1,1,r^2,r^2\sin^2\theta)$. The function $\Phi (r)$ can be interpreted as the gravitational potential. For typical stellar cores, this can be modelled as having the form $\Phi (r)=-\lambda r^2(4-3r)$. The dimensionless perturbing parameter $\lambda$ is the same as the scalar-field coupling parameter since we assumed that the two sources of perturbations have comparable strengths.

In the orthonormal frame,the operator in Eq.(20) becomes
\begin{eqnarray}
\square^{(1)}\Psi =\frac{\Phi}{r^2}\frac{\partial}{\partial r}\bigg(r^2\frac{\partial\Psi}{\partial r}\bigg)+\frac{1}{r^2}\frac{\partial}{\partial r}\bigg(r^2\Phi\frac{\partial\Psi}{\partial r}\bigg).
\end{eqnarray}

As discussed previously, the superfluid does not have a simple ball geometry. A more complete determination of its shape involves solving the metric and the order parameter from the Tollman-Oppenheimer-Volkov and Hartree equations simultaneously. In this work, our goal is only to investigate the effect of a fixed metric background on the superfluid and so we will not pursue this type of complicated calculation. Instead, we imagine a mathematical sphere of radius unity inside the superfluid core. We then assume that the value of the order parameter is known on this boundary. For concreteness, we consider two examples familiar in electromagnetic theory\cite{Griffiths99, Jackson99}.
\subsection*{Example 1}
As our first example, consider the boundary condition
\begin{equation}
\Psi^{(0)}(r=1,\theta ,\phi)=\left\{
\begin{array}{lr}
+1 &,0\leq \theta <\frac{\pi}{2}\\
-1 &,\frac{\pi}{2}<\theta\leq \pi
\end{array}
\right.
\end{equation}

The zeroth-order solution is given by
\begin{eqnarray}
\Psi^{(0)}(r,\theta)=\sum_{l=o}^{\infty}A_lr^lP_l(\cos\theta)
\end{eqnarray}
where
\begin{eqnarray}
A_l=\frac{2l+1}{2}\bigg[\int_0^1P_l(x)dx-\int_{-1}^0P_l(x)dx\bigg]
\end{eqnarray}
and $P_l(x)$ is a Legendre polynomial.

\begin{figure}[h!]
  \centering
     \includegraphics[width=8.25cm,height=6cm]{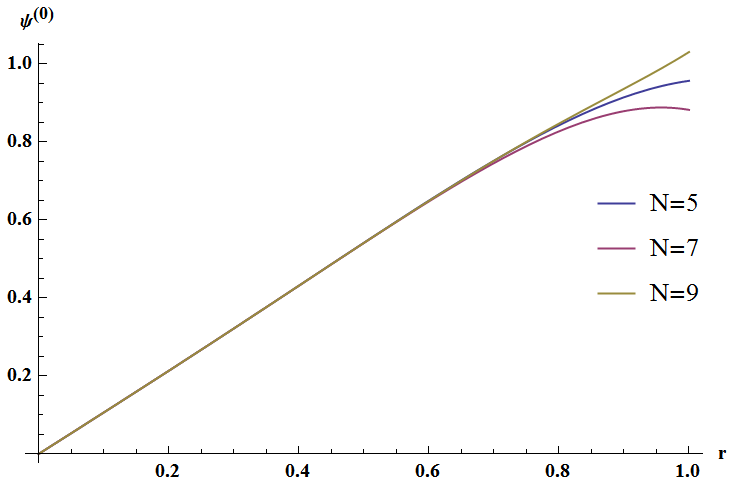}
    \caption[]{(Color Online) Plot of $\Psi^{(0)}$ vs $r$ for different truncations $N=5$, $N=7$, and $N=9$, of the series solution for $\Psi^{(0)}$. The value $\theta=\pi/4$ was used.}
\end{figure}

Shown in Figure 1 is the plot of Eq.(33) using different truncations $N=5$, $N=7$, and $N=9$ in the summation $\sum_{l=0}^N\cdot\cdot\cdot$ for $\theta=\frac{\pi}{4}$. We see that there is convergence for $r<0.8$. This convergence is also seen for other values of $\theta$. In our subsequent calculations, we retained terms in the summations up to $N=5$, keeping in mind that our results are reliable up to radius $r\sim 0.8$.

\begin{figure}[h!]
  \centering
     \includegraphics[width=8.25cm,height=6cm]{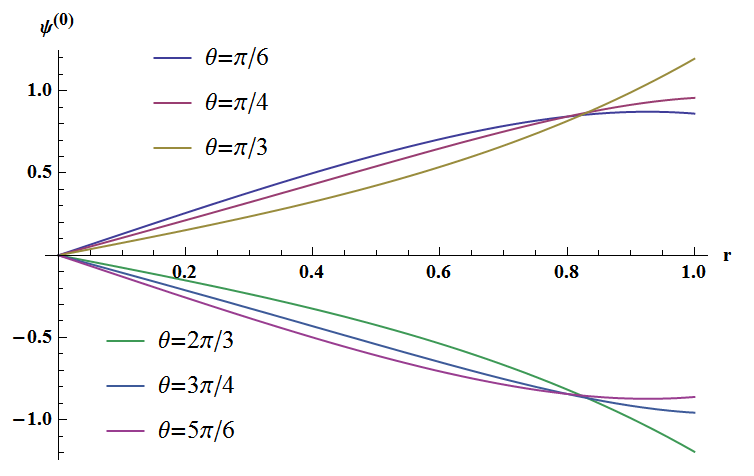}
    \caption[]{(Color Online) Plot of $\Psi^{(0)}$ vs $r$ for different values of $\theta$.}
\end{figure}

Figure 2 shows the plot of the zeroth-order contribution to the order parameter $\Psi^{(0)}$ versus radius $r$. We see that for this boundary condition (32), $\Psi^{(0)}$ vanishes at the center $r=0$.

\begin{figure}[h!]
  \centering
     \includegraphics[width=8.25cm,height=6cm]{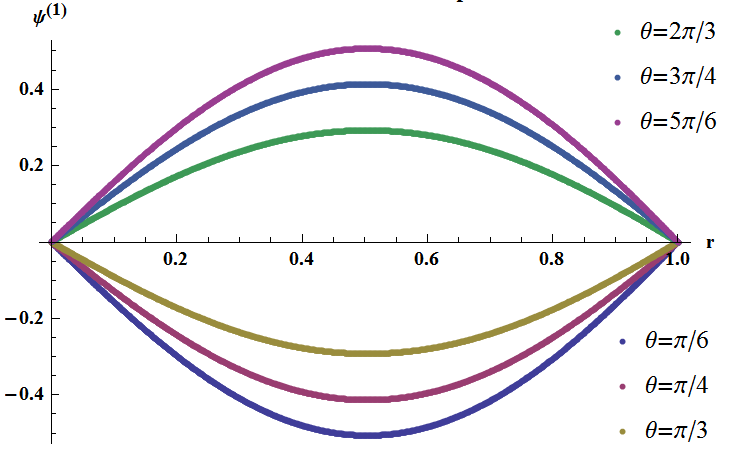}
    \caption[]{(Color Online) Plot of $\Psi^{(1)}$ vs $r$ for different values of $\theta$.}
\end{figure}

The plot of the gravitational contribution to the first-order correction of the order parameter is shown in Figure 3 for various values of $\theta$. We note from Figures 2 and 3 that for $0\leq \theta <\pi/2$, the zeroth order is positive $\Psi^{(0)}\geq 0$ while the gravitational correction has the opposite sign $\Psi^{(1)}\leq 0$. For $\pi/2\leq\theta \leq\pi$, $\Psi^{(0)}\leq 0$ while $\Psi^{(1)}\geq 0$. That is, for the boundary condition considered, the first-order gravitational  correction $\Psi^{(1)}$ and the zeroth order $\Psi^{(0)}$ have opposite signs. Hence, in this case, gravity tends to destroy the superfluidity. Note also that in the middle part $r\sim 0.5$, the correction is greatest $\Psi^{(1)}\sim 0.5$. Since $\Psi=\Psi^{(0)}+\lambda\Psi^{(1)}$, this means that the correction can reach values up to $\sim 0.5\lambda$ even with our assumption of weak gravity. This example shows that gravity can have a significant effect on the formation of superfluids inside compact stars.

\subsection*{Example 2}

\begin{figure}[h!]
  \centering
     \includegraphics[width=8.25cm,height=6cm]{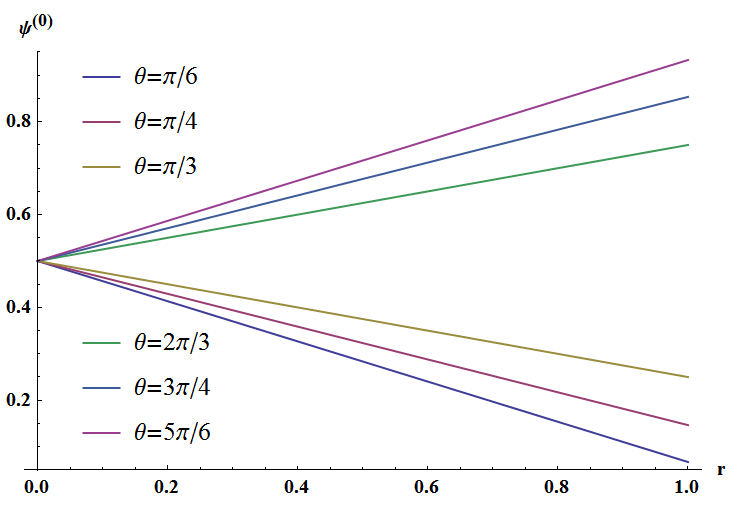}
    \caption[]{(Color Online) Plot of $\Psi^{(0)}$ vs $r$ for different values of $\theta$.}
\end{figure}

For our second example, we consider other boundary condition 
\begin{eqnarray}
\Psi^{(0)}(1,\theta)=\sin^2\frac{\theta}{2}.
\end{eqnarray}

The zeroth-order solution is simply
\begin{eqnarray}
\Psi^{(0)}(r,\theta)=\frac{1}{2}(1-r\cos\theta).
\end{eqnarray}
The plots for various $\theta$ are shown in Figure 4.

\begin{figure}[h!]
  \centering
     \includegraphics[width=8.25cm,height=6cm]{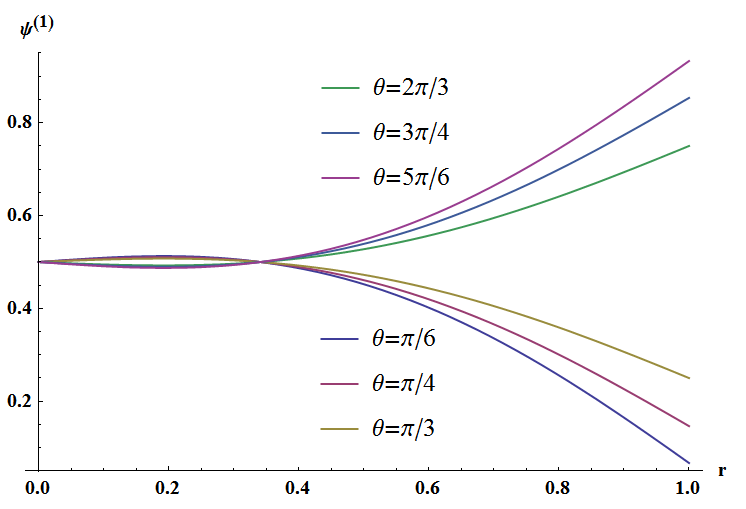}
    \caption[]{(Color Online) Plot of $\Psi^{(1)}$ vs $r$ for different values of $\theta$.}
\end{figure}

The gravitational contribution to the first-order correction is plotted in Figure 5 for various values of $\theta$. For this specific boundary condition, gravity enhances the formation of the superfluid. We note that the gravitational correction can reach values up to $\sim 0.7\lambda$, which shows again that gravity has a significant effect on the formation of the superfluid.

We have seen in this section that gravity can have significant and non-trivial effects on the formation of relativistic scalar-field superfluids. In the next section, we consider relativistic superfluids formed by fermions pairs.

\section{III. The effective action}
\label{sec:effact}
\subsection{Scalar boson-mediated interaction}
Let us consider a system of fermions interacting with a massless scalar bosons via a Yukawa term in curved spacetime. Its generating functional can be written as
\begin{equation}
\mathcal{Z}=\mathcal{N}\int\mathcal{D}\bar{\psi}\mathcal{D}\psi\mathcal{D}\phi
\exp \left\{iS[\bar{\psi},\psi,\phi]\right\}
\end{equation}
where the action is given by the integral of the Lagrangian density
\begin{eqnarray}
S[\bar{\psi},\psi,\phi]&=&\int d^4x\sqrt{-g(x)}\bigg[\bar{\psi}(i\gamma^{\nu}\nabla_{\nu}-m+\mu\gamma_{\hat{0}})\psi\nonumber\\
{}&{}&-\lambda\bar{\psi}\psi\phi +\frac{1}{2}g^{\nu\rho}\partial_{\nu}\phi\partial_{\rho}\phi\bigg].
\end{eqnarray}
Here $g(x)\equiv\det g_{\mu\nu}(x)$ and $\nabla_{\nu}\equiv \partial_{\nu}+\Gamma_{\nu}$, where $\Gamma_{\nu}$ is the connection appropriate for a fermion field. Indices without hats are general coordinate indices, $\mu$ is the chemical potential, and $\lambda$ is the Yukawa coupling. The gamma symbol with a hatted index $\gamma_{\hat{0}}$ is the usual Dirac gamma matrix in flat spacetime. It is not converted into a general Dirac gamma matrix in curved spacetime since its role is only to rearrange the components of the spinor field to get the appropriate number operator for the chemical potential term. 

The scalar-boson degree of freedom can be integrated by rewriting its kinetic term using the identities 
\begin{eqnarray}
\Box \phi &\equiv &g^{\mu\nu}\nabla_{\mu}\nabla_{\nu}\phi\\
&=&(-g)^{-1/2}\partial_{\mu}\left[(-g)^{1/2}g^{\mu\nu}\partial_{\nu}\phi\right],\\ \Gamma^{\mu}_{\mu\alpha}&=&\frac{1}{\sqrt{-g}}\partial_{\alpha}\sqrt{-g},
\end{eqnarray}
and using the Gauss-Stokes theorem. The value of the scalar field is taken to be zero at infinity. 

The Lagrangian of the scalar field, including its coupling with the fermion field via the Yukawa term, then becomes
\begin{equation}
\mathcal{L}=-\frac{1}{2}(-g)^{1/2}\phi g^{\mu\nu}\nabla_{\mu}\nabla_{\nu}\phi-(-g)^{1/2}j\phi
\end{equation}
where $j(x)\equiv \lambda\bar{\psi}(x)\psi (x)$ is the fermion current.

The functions $\phi(x)$ and $j(x)$ can be expanded in terms of some convenient set of complete functions $\phi_n(x)$ which are chosen to be the eigenfunctions of the eigenvalue equation
\begin{equation}
g^{\mu\nu}\nabla_{\mu}\nabla_{\nu}\phi_n=A_n\phi_n
\end{equation}
and obeying the orthonormal condition
\begin{equation}
\int d^4x\sqrt{-g}\phi_n\phi_m=\delta_{n,m}.
\end{equation}

The path integral measure over the $\phi$ field becomes
\begin{equation}
\mathcal{D}\phi\rightarrow \prod_n\int_{-\infty}^{\infty}da_n.
\end{equation}

After completing the square by shifting the fields, the path integral becomes
\begin{eqnarray}
\int\mathcal{D}\phi e^{iS}&=&\left(\prod_n\int_{-\infty}^{\infty}d\tilde{a}_n e^{-\frac{1}{2}A_n\tilde{a}^2_n}\right)\nonumber\\
{}&{}&\times\exp \left(\frac{i}{2}\sum_n b_nA^{-1}_nb_n\right)
\end{eqnarray}
where $\tilde{a}_n$ and $b_n$ are the coefficients in the expansion of $\phi(x)$ and $j(x)$ in terms of $\phi_n$, respectively.

The first factor will cancel in the calculation of the correlation functions and hence has no contribution in the calculation of physical quantities. The second factor gives the effective action. This can be rewritten back into the real space by inverting the expansion formula for $j(x)$: $b_n=\int d^4x\sqrt{-g}\phi_n(x)j(x)$. The effective four-fermion interaction then becomes
\begin{widetext}
\begin{equation}
S_{eff. int}=\frac{\lambda^2}{2}\int d^4x\sqrt{-g(x)}d^4y\sqrt{-g(y)}\bar{\psi}(x)\psi(x)D(x,y)\bar{\psi}(y)\psi(y)
\end{equation}
\end{widetext}
where
\begin{equation}
D(x,y)\equiv \sum_n\frac{[\phi_n(x)][\phi_n(y)]}{A_n}.
\end{equation}

Equation (47) gives the general effective four-fermion interaction. It has a similar form to that of flat spacetime case\cite{Pisarski99}, except that the factor $D(x,y)$, which is the scalar field propagator in the Minkowski case, is now to be evaluated by summing the terms involving the eigenfunctions and eigenvalues of the eigenvalue equation (43). The path integral involving an interaction of the form Eq.(47) is difficult, if not impossible, to evaluate. To make progress, a mean-field approach or a Hubbard-Stratonovich transformation is usually employed. This is what we do in Section IV. For the meantime, let us discuss briefly a more physically motivated extension of the case treated here, which is the fermion interaction mediated by gauge vector bosons. In various quark pairings, for example, the mediators are the gauge vector bosons of the SU(3) gauge field.

\subsection{Gauge vector boson-mediated interaction} 
In curved spacetime, the action for fermion and gluon fields can be written as
\begin{widetext}
\begin{eqnarray}
S&=&\int\mbox{d}^4x\sqrt{-g(x)}\int\mbox{d}^4y\sqrt{-g(y)}\left[\bar{\psi}(x)G^{-1}_0(x,y)\psi(y)-\frac{1}{2}\sum_{c=1}^8A^c_{\mu}(x)D^{-1}_0(x,y)^{\mu\nu}A^c_{\nu}(y)\right]\nonumber\\
{}&{}&-g\int\mbox{d}^4x\sqrt{-g(x)}\sum_{c=1}^8\bar{\psi}(x)\gamma^{\mu}(x)A^c_{\mu}(x)T^c_3\psi(x)+S_{gluon-int.}
\end{eqnarray}
\end{widetext}
where $A^c_{\mu}$ are the gluon fields, $T^c_3$ are the generators of the fundamental representation of SU(3), and $S_{gluon-int}$ are the gluon interactions arising from the non-abelian nature of SU(3).

The inverse Feynman free-fermion propagator is given by 
\begin{equation}
G^{-1}_0(x,y)=(\det V)(i\gamma^{\mu}(x)\nabla_{\mu}-\mu\gamma^{\hat{0}}-m)\delta^{(4)}(x-y)
\end{equation}
where $V_{\hat{\alpha}}^{\mu}(x)$ is the vierbein and $\gamma^{\mu}(x)$ is the generalization of the gamma matrices in curved spacetime. They can be written in terms of the usual gamma matrices in flat spacetime by projecting via the vierbein $\gamma^{\mu}(x)=V^{\mu}_{\hat{\alpha}}(x)\gamma^{\hat{\alpha}}$. As before, indices with hats denote local frame; while, those without hats denote general coordinate component. The covariant derivative is given by $\nabla_{\hat{\alpha}}=V^{\mu}_{\hat{\alpha}}(\partial_{\mu}+\Gamma_{\mu})$, where $\Gamma$ is an appropriate connection for the field in which the derivative acts.

The propagator for the gluon, on the other hand, satisfies
\begin{eqnarray}
[g_{\mu\rho}(x)\square_x &+&R_{\mu\rho}(x)-(1-\xi^{-1})\nabla^x_{\mu}\nabla^x_{\rho}]D_0(x,y)^{\rho\nu}\nonumber\\
&=&[-g(x)]^{-1/2}\delta^{\nu}_{\;\mu}\delta^{(4)}(x-y)
\end{eqnarray}
where $\xi$ comes from a gauge-fixing term.

If the pure gluon interactions are neglected, the gluon fields can be formally integrated-out to yield the effective action functional
\begin{equation}
\mathcal{Z}\sim (\det D^{-1}_0)^{-1/2}\int\mathcal{D}\bar{\psi}\mathcal{D}\psi e^{i\tilde{S}[\bar{\psi},\psi]}
\end{equation}
where
\begin{widetext}
\begin{eqnarray}
\tilde{S}[\bar{\psi},\psi]=\int d^4x\sqrt{-g(x)}d^4y\sqrt{-g(y)}\left[\bar{\psi}(x)G^{-1}_0(x,y)\psi(y)+\frac{g^2}{2}\bar{\psi}(x)\gamma_{\mu}(x)T^c_3\psi(x)D^{\mu\nu}_0(x,y)\bar{\psi}(y)\gamma_{\nu}(y)T^c_3\psi(y)\right].
\end{eqnarray}
\end{widetext}
For sufficiently weak couplings, such as in color-flavor locked phase of quark matter, $S_{gluon-int}$ can be accounted by perturbative expansion. Its effect is to renormalized the propagator $D^{\mu\nu}_0(x,y)$ via self-energy corrections. For strongly-coupled case, one has to resort to lattice gauge and numerical calculations.
Equation (53) is the general effective four-fermion interaction in curved spacetime obtained from the gauge vector boson-mediated fermion interaction. 

In the next section, we go back to the simpler scalar mediated case, equation (47), and apply mean-field ansatz to simplify the effective action. The methods that will be used should also be applicable to the gauge vector boson-mediated case by a straightforward extension.

\section{IV. Mean-field pairing ansatz}
\subsection{Antifermion-fermion pairing}
\label{sec:antif}
If there is an antifermion-fermion pairing, then the non-vanishing expectation value $\langle j(x)\rangle\equiv \langle\lambda\bar{\psi}(x)\psi(x)\rangle\neq 0$ can be chosen as the order parameter. The fluctuations about this mean value $\rho (x)\equiv j(x)-\langle j(x)\rangle$ is then introduced.
It is assumed that this fluctuation is small so that equation (47) can be expanded only up to the terms that are linear in $\rho (x)$. The function $\rho (x)$ is then traded back with $j(x)$ by using $j(x)=\rho (x)+\langle j(x)\rangle$ yielding
\begin{widetext}
\begin{eqnarray}
S_{eff.int}=\frac{1}{2}\int d^4x\sqrt{-g(x)}d^4y\sqrt{-g(y)}[2j(x)D(x,y)\langle j(y)\rangle -\langle j(x)\rangle D(x,y)\langle j(y)\rangle].
\end{eqnarray}
\end{widetext}

The full action is therefore
\begin{equation}
S=\int d^4x\sqrt{-g(x)}\bar{\psi}(x)\left[i\gamma^{\alpha}\nabla_{\alpha}-m+\mu\gamma_{\hat{0}}+\lambda^2 F(x)\right]\psi (x)
\end{equation}
where
\begin{equation}
F(x)\equiv \int d^4y\sqrt{-g(y)}D(x,y)\langle\bar{\psi}(y)\psi(y)\rangle.
\end{equation}

The eigenfunctions of the eigenvalue equation
\begin{equation}
[i\gamma^{\alpha}\nabla_{\alpha}-m+\mu\gamma_{\hat{0}}+\lambda^2 F(x)]\pi_n(x)=B_n(\mu,\lambda)\pi_n(x),
\end{equation}
which obeys the orthonormal condition
\begin{equation}
\int d^4x\sqrt{-g(x)}\bar{\pi}_n(x)\pi_m(x)=\delta_{m,n}
\end{equation}
are now used as an orthonormal basis. Note that the eigenvalues $B_n(\mu,\lambda)$ are functions of the chemical potential $\mu$ and the Yukawa coupling $\lambda$.

The fields are expanded as
\begin{equation}
\psi(x)=\sum_nc_n\pi_n(x).
\end{equation}
Here, the eigenfunctions $\pi_n(x)$ are four-component c-number functions while the coefficients $c_n$ are Grassmann variables.

The action can now be written as
\begin{eqnarray}
S[\bar{\psi},\psi]=\sum_nc^{\dagger}_nB_n(\mu ,\lambda)c_n.
\end{eqnarray}

This can be rewritten back into real space by inverting equation (59). This yields
\begin{equation}
S[\bar{\psi},\psi]=\int d^4x\sqrt{-g(x)}\int d^4y\sqrt{-g(y)}\bar{\psi}(x)W(x,y)\psi(y)
\end{equation}
where
\begin{equation}
W(x,y)\equiv \sum_n\pi_n(x)B_n(\mu ,\lambda)\bar{\pi}_n(y)
\end{equation}
is a $4\times 4$ Dirac-spinor matrix.

This mean-field effective action is now quadratic in the fermion field $\psi$. Its path integral, including the various correlation functions of interest, can now be evaluated. Note that the crucial ingredient to do this is the evaluation of the eigenfunctions $\pi_n(x)$ and eigenvalues $B_n(\mu,\lambda)$ of the eigenvalue equation (57).

The self-consistency equation that must be satisfied is
\begin{eqnarray}
\langle \bar{\psi}(z)\psi(z)\rangle &\equiv & \mbox{Tr}\big\{W(z,z)\big\}\\
&=&\sum_n\mbox{Tr}\bigg\{\pi_n(z)B_n(\mu ,\lambda)\bar{\pi}_n(z)\bigg\}\\
&=&\sum_n\bar{\pi}_n(z)\pi_n(z)B_n(\mu ,\lambda).
\end{eqnarray}
Note that the last two expressions in the right involve the eigenfunctions $\pi_n(x)$ and eigenvalues $B_n(\mu,\lambda)$ that are to be determined from the eigenvalue equation (57), which, in turn, contains the order parameter $\langle \bar{\psi}\psi\rangle$ via the function $F(x)$ defined in equation (56). Hence, equation (65) is to be determined self-consistently.

We can get the explicit analytical expression for the gravitational correction to the order parameter if we assume that the gravity is sufficiently weak so that the metric can be written as a sum of the flat metric and a perturbation $g_{\mu\nu}=\eta_{\mu\nu}+h_{\mu\nu}$. If we expand the eigenfunctions and eigenvalues perturbatively, $\pi_n(z)=\pi^{(0)}_n(z)+\pi^{(1)}_n(z)\cdot\cdot\cdot$ and $B_n=B^{(0)}_n+B^{(0)}_n\cdot\cdot\cdot$, then from (65), we have the zeroth and first order contributions to the order parameter
\begin{eqnarray}
\langle\bar{\psi}(z)\psi(z)\rangle^{(0)}=\sum_n\bar{\pi}^{(0)}_n\pi^{(0)}_nB^{(0)}_n\\
\end{eqnarray}
and
\begin{eqnarray}
\langle\bar{\psi}(z)\psi(z)\rangle^{(1)}&=&\sum_n\big(\bar{\pi}^{(1)}_n\pi^{(0)}_nB^{(0)}_n+\bar{\pi}^{(1)}_n\pi^{(0)}_nB^{(0)}_n\nonumber\\
&{}&{}+\bar{\pi}^{(0)}_n\pi^{(0)}_nB^{(1)}_n\big),
\end{eqnarray}
respectively.

The first order correction to the eigenfunctions $\pi_n^{(1)}$ and eigenvalues $B_n^{(1)}$ can be obtained by using the standard perturbation theory to solve Equation (57). We assumed that the gravitational perturbation has comparable strength to the weak dimensionless Yukawa coupling $\lambda$. Note that the self-consistent term $\lambda^2F(x)$ is second order in $\lambda$. Hence, the first order corrections come purely from the gravitational contribution.

The first-order correction to the eigenvalues is given by
\begin{eqnarray}
B^{(1)}_n&=&\int d^4x\sqrt{-g(x)}\bar{\pi}^{(0)}_ni\gamma^{\hat{\alpha}}\big[V_{\hat{\alpha}}^{\mu(0)}\Gamma^{(1)}_{\mu}\nonumber\\
{}&{}&+V^{\mu(1)}_{\hat{\alpha}}(\partial_{\mu}+\Gamma^{(0)}_{\mu})\big]\pi^{(0)}_n;
\end{eqnarray}
while the first-order correction to the eigenfunctions is
\begin{eqnarray}
\pi^{(1)}_n=\sum_{m\neq n}c_m\pi^{(0)}_m
\end{eqnarray}
with
\begin{eqnarray}
c_m&=&\frac{1}{B^{(0)}_n-B^{(0)}_m}\int d^4x\sqrt{-g(x)}\bar{\pi}^{(0)}_mi\gamma^{\hat{\alpha}}\big[V^{\mu(0)}_{\hat{\alpha}}\Gamma^{(1)}_{\mu}\nonumber\\
{}&{}&+V^{\mu(1)}_{\hat{\alpha}}(\partial_{\mu}+\Gamma^{(0)}_{\mu})\big]\pi^{(0)}_n.
\end{eqnarray}

In the equations above, we have expanded the vierbein and the connection in powers of the strength of the gravitational perturbation $\lambda$: $V^{\mu}_{\hat{\alpha}}=V^{\mu(0)}_{\hat{\alpha}}+V^{\mu(1)}_{\hat{\alpha}}+\cdot\cdot\cdot$ and $\Gamma_{\mu}=\Gamma^{(0)}_{\mu}+\Gamma^{(1)}_{\mu}+\cdot\cdot\cdot$.

The type of pairing considered here typically occurs in chiral symmetry breaking phase of QCD which typically happens in the low-density regime, although they might be relevant in the formation of meson condensates inside star cores. Let us now consider the type of pairing which is suggested to occur in the high-density regime. This is the fermion-fermion pairing. Examples of this type of pairing are the two-color superconducting state and the color-flavor locked phase in dense quark matter. They can possibly exist at the core of neutron stars or in quark stars. We consider this in the next discussion.

\subsection{Fermion-fermion pairing}
\label{sec:antif}
In this subsection, we consider the fermion-fermion pairing ansatz. Since this type of pairing is physically well motivated, due to the possibility of color-superconducting and color-flavor locked phases in quark matter, we will derive additional quantities: the Euclidean effective action and the matrix elements of the heat kernel operator, which are very important in curved-spacetime quantum field theory calculations.

We introduce the charge-conjugate spinor $\psi_C\equiv \mathcal{C}\bar{\psi}^T(x)$, $\bar{\psi}_C=\psi^T\mathcal{C}$, $\psi=\mathcal{C}\bar{\psi}^T_C$, and $\bar{\psi}=\psi^T_C\mathcal{C}$. The charge conjugation simply interchanges the particle with the antiparticle so it is taken to be numerically equal to $\mathcal{C}=i\gamma^{\hat{2}}\gamma^{\hat{0}}$, where the gammas are the usual Lorentz-indexed Dirac gamma matrices. The charge conjugation operator obeys the usual $\mathcal{C}=-\mathcal{C}^{-1}=-\mathcal{C}^T=-\mathcal{C}^{\dagger}$.
The four-fermion interaction term in equation (47) can then be rewritten as
\begin{eqnarray}
\bar{\psi}(x)\psi(x)\bar{\psi}(y)\psi(y)
&=&-\frac{1}{2}\mbox{Tr}\big\{[\psi(y)\bar{\psi}_C(x)][\psi_C(x)\bar{\psi}(y)]\nonumber\\
{}&{}&+h.c.\big\}.
\end{eqnarray}

Introduce the matrix $J(x,y)\equiv\psi_C(x)\bar{\psi}(y)$ and its conjugate $J^{\dagger}(y,x)=\beta\psi(y)\bar{\psi}_C(x)\beta$
so that (72) is written as
\begin{equation}
\bar{\psi}(x)\psi(x)\bar{\psi}(y)\psi(y)=-\frac{1}{2}\mbox{Tr}\bigg\{\beta J^{\dagger}(y,x)\beta J(x,y)+h.c.\bigg\}.
\end{equation}

The fermion-fermion pairing order parameter can then be taken as $\langle J(x,y)\rangle $ and the fluctuations about this matrix mean-field value is now $\rho (x,y)\equiv J(x,y)-\langle J(x,y)\rangle$. This fluctuation is assumed to be sufficiently small so that only the terms linear in $\rho (x,y)$ have a significant contribution. The interaction term can then be rewritten as
\begin{eqnarray}
\bar{\psi}(x)\psi(x)\bar{\psi}(y)\psi(y)&=&\bar{\psi}_C(x)\langle\psi_C(x)\bar{\psi}(y)\rangle\psi (y)\nonumber\\
{}&{}&+\bar{\psi}(y)\langle\psi(y)\bar{\psi}_C(x)\rangle\psi_C(x).
\end{eqnarray}

It is convenient to introduce Nambu's notation
\begin{eqnarray}
\Psi (x)=
\begin{pmatrix}
\psi(x)\\
\psi_C(x)
\end{pmatrix}
\end{eqnarray}
and rewrite the full action as
\begin{widetext}
\begin{eqnarray}
S=\int d^4x\sqrt{-g(x)}\int d^4y\sqrt{-g(y)}
\begin{pmatrix}
\bar{\psi}(x), & \bar{\psi}_C(x)
\end{pmatrix}
\begin{pmatrix}
\hat{W}^+_F(x,y) & \Delta^{\dagger}(x,y)\\
\Delta (x,y) & \hat{W}^-_F(x,y)
\end{pmatrix}
\begin{pmatrix}
\psi (y)\\
\psi_C(y)
\end{pmatrix}
\end{eqnarray}
\end{widetext}
where $\hat{W}^{\pm}_F(x,y)\equiv \frac{1}{2}(i\gamma^{\alpha}\nabla_{\alpha}-m\pm \mu\gamma_{\hat{0}})\delta^{(4)}(x-y)$ and the gap variables are given by $\Delta = \frac{\lambda^2}{2}D(x,y)\langle\psi(x)\bar{\psi}_C(y)\rangle $ and $\Delta^{\dagger} = \frac{\lambda^2}{2}D(x,y)\langle\psi_C(x)\bar{\psi}(y)\rangle$.

This is the general effective action for a fermion-fermion pairing mean-field ansatz mediated by a scalar boson. We want to further simplify this as possible. First, consider the diagonal term (rewritten back in terms of $\psi$ and $\bar{\psi}$):
\begin{eqnarray}
\int d^4x\sqrt{-g}\bar{\psi}(x)(i\gamma^{\alpha}\nabla_{\alpha}-m+\mu\gamma_{\hat{0}})\psi(x).
\end{eqnarray}

We expand the fields $\psi$ and $\bar{\psi}$ in terms of the eigenfunctions $\pi_n$ of the eigenvalue equation
\begin{eqnarray}
(i\gamma^{\alpha}\nabla_{\alpha}-m+\mu\gamma_{\hat{0}})\pi_n(x)=\lambda_n\pi_n(x)
\end{eqnarray}
that obey the orthonormal condition
\begin{eqnarray}
\int d^4x\sqrt{-g}\bar{\pi}_m(x)\pi_n(x)=\delta_{m,n}.
\end{eqnarray}

We have
\begin{eqnarray}
\psi(x)&=&\sum_na_n\pi_n(x)\\
\bar{\psi}(x)&=&\sum_na_n^{\dagger}\bar{\pi}_n(x)
\end{eqnarray}
where $\pi_n$ and $\bar{\pi}_n$ are four-component c-numbers and $a$ and $a^{\dagger}$ are Grassmann variables.

The expression (77) becomes
\begin{eqnarray}
\int d^4x\sqrt{-g}\bar{\psi}(x)(i\gamma^{\alpha}\nabla_{\alpha}-m+\mu\gamma_{\hat{0}})\psi(x)=\sum_n\lambda_na^{\dagger}_na_n.
\end{eqnarray}

We rewrite this back in terms of fields $\psi$ and $\bar{\psi}$ by inverting Equations (80) and (81)
\begin{eqnarray}
S_{diag}=\int d^4x\sqrt{-g(x)}d^4y\sqrt{-g(y)}\bar{\psi}(x)G(x,y)\psi(y)
\end{eqnarray}
where
\begin{eqnarray}
G(x,y)\equiv \sum_n\lambda_n\pi_n(x)\bar{\pi}_n(y).
\end{eqnarray}

The action can now be written as
\begin{widetext}
\begin{eqnarray}
S=\int d^4x\sqrt{-g(x)}d^4y\sqrt{-g(y)}
\begin{pmatrix}
\bar{\psi}(x), & \bar{\psi}_C(x)
\end{pmatrix}
\begin{pmatrix}
\frac{1}{2}G(x,y) & \Delta^{\dagger}(x,y)\\
\Delta (x,y) & \frac{1}{2}G_C(x,y)
\end{pmatrix}
\begin{pmatrix}
\psi (y)\\
\psi_C(y)
\end{pmatrix}
\end{eqnarray}
\end{widetext}
where $G_C(x,y)\equiv C\gamma_0G^*(x,y)\gamma_0C$.

Unlike equation (76) which contains derivative operators via the diagonal terms $\hat{W}_F^{\pm}(x,y)$, the matrix in the equation above are pure numbers. We can now diagonalize it using a unitary transformation
\begin{eqnarray}
\begin{pmatrix}
\psi\\
\psi_C
\end{pmatrix}
=
\begin{pmatrix}
e^{i\phi}\cos \theta & e^{i\phi}\sin \theta \\
-e^{-i\phi}\sin \theta & e^{-i\phi}\cos\theta
\end{pmatrix}
\begin{pmatrix}
\chi\\
\xi
\end{pmatrix}
\end{eqnarray}
with 
\begin{eqnarray}
e^{2i\phi}&=&\sqrt{\frac{\Delta^{\dagger}}{\Delta}}\\
\tan 2\theta &=&\frac{2|\Delta|}{G_C-G}.
\end{eqnarray}

This yields
\begin{eqnarray}
S&=&\int d^4x\sqrt{-g(x)}d^4y\sqrt{-g(y)}
\begin{pmatrix}
\bar{\chi}(x), & \bar{\xi}(x)
\end{pmatrix}\nonumber\\
&{}&{}\times
\begin{pmatrix}
D(x,y)_- & 0\\
0 & D(x,y)_+
\end{pmatrix}
\begin{pmatrix}
\chi(y) \\
\xi(y)
\end{pmatrix}
\end{eqnarray}
where
\begin{equation}
D(x,y)_{\pm}\equiv \frac{1}{4}(G+G_C)\pm \frac{(G_C-G)^2+8|\Delta |^2}{4\sqrt{4|\Delta |^2+(G_C-G)^2}}.
\end{equation}

In curved spacetime QFT calculations, it is usually more convenient to consider the Euclidean effective action. To do this, the Wick rotations $x^0\rightarrow ix^0$ and $\gamma^0\rightarrow i\gamma^0$ are performed. This is done for the contravariant time components both for general and local coordinates. The resulting Euclidean action is 
\begin{eqnarray}
S_E=i\int d^4x_E\sqrt{g^E(x)}
\begin{pmatrix}
\bar{\chi}, & \bar{\xi}
\end{pmatrix}
\begin{pmatrix}
D_- & 0\\
0 & D_+
\end{pmatrix}
\begin{pmatrix}
\chi \\
\xi
\end{pmatrix}
\end{eqnarray}
where $g^E_{\mu\nu}(x)\equiv -g_{\mu\nu}(it,\vec{x})$. The form of the operators $\hat{d}\pm\hat{g}$ are preserved.

The Euclidean effective action $\Gamma_E[g^E_{\mu\nu}]$ is then defined as
\begin{equation}
e^{-\Gamma_E[g_{\mu\nu}]}=\int\mathcal{D}\bar{\Psi}\mathcal{D}\Psi
e^{-S_E[\bar{\Psi},\Psi,g_{\mu\nu}]}
\end{equation}
where
\begin{eqnarray}
\Psi (x)\equiv
\begin{pmatrix}
\chi(x)\\
\xi(x)
\end{pmatrix}
\end{eqnarray}

The path integral can be evaluated yielding the effective action
\begin{equation}
\Gamma_E[g_{\mu\nu}]=-\mbox{ln}\det\big[D_-D_+\big].
\end{equation}

This analytical expression of the euclidean effective action for the fermion-fermion pairing, expressed as a functional determinant is one of the main result in this section. This is a very useful starting point in calculating various quantities like correlation functions. 

Most quantities of interest in curved-spacetime quantum field theory, like the expectation value of the energy-momentum tensor, are calculated via the heat kernel method\cite{Mukhanov07}. An important quantity in this approach is the heat kernel operator $\hat{K}$. Let us therefore derived an analytical expression for the matrix elements of this quantity.

The heat kernel satisfies the differential equation
\begin{equation}
\frac{d\hat{K}(\tau)}{d\tau}=-\hat{O}\hat{K}(\tau)
\end{equation}
with the initial condition $\hat{K}(0)=1$.

In terms of some basis $\{|x\rangle\}$ the matrix elements of the operator $\hat{O}$ is given by(see Appendix)
\begin{equation}
\langle x|\hat{O}_{\pm}|x'\rangle =g^{1/4}(x)D(x)_{\pm}[g^{-1/4}(x)\delta (x-x')].
\end{equation}

If the gravitational field is a sufficiently weak perturbation from the flat metric $g_{\mu\nu}=\eta_{\mu\nu}+h_{\mu\nu}$, then one can expand the various quantities like the vierbein $V^{\mu}_{\hat{\alpha}}=V^{(0)\mu}_{\hat{\alpha}}+V^{(1)\mu}_{\hat{\alpha}}\cdot\cdot\cdot$, the connection $\Gamma_{\mu}=\Gamma^{(0)}_{\mu}+\Gamma^{(1)}_{\mu}\cdot\cdot\cdot$ and the various quantities $D(x)_{\pm}=D(x)_{\pm}^{(0)}+D(x)_{\pm}^{(1)}+\cdot\cdot\cdot$ and $\langle x|\hat{O}_{\pm}|x'\rangle = \langle x|\hat{O}_{\pm}|x'\rangle^{(0)}+\langle x|\hat{O}_{\pm}|x'\rangle^{(1)}+\cdot\cdot\cdot$.

The matrix elements of the zeroth and first order expressions for the operator $\hat{O}_{\pm}$ on the other hand are given by
\begin{widetext}
\begin{eqnarray}
\langle x|\hat{O}_{\pm}|x'\rangle^{(0)}&=&D(x)_{\pm}^{(0)}\delta (x-x')\\
\langle x|\hat{O}_{\pm}|x'\rangle^{(1)}&=&D(x)_{\pm}^{(0)}[\frac{1}{4}\delta_{\rho\sigma}h^{\rho\sigma}\delta(x-x')]-\frac{1}{4}\delta_{\rho\sigma}h^{\rho\sigma}D(x)_{\pm}^{(0)}\delta(x-x')+D(x)_{\pm}^{(1)}\delta(x-x').
\end{eqnarray}
\end{widetext}

The heat kernel can similarly be expanded $\hat{K}(\tau)=\hat{K}_0(\tau)+\hat{K}_1(\tau)+\cdot\cdot\cdot$. If one knows the matrix elements of the zeroth order, one can then solve the elements of the first order:
\begin{equation}
\hat{K}_1(\tau)=-\int_0^{\tau}d\tau'\hat{K}_0(\tau-\tau')\hat{O}^{(1)}\hat{K}_0(\tau').
\end{equation}

The zeroth-order solution to equation (95) is 
\begin{eqnarray}
\langle x|\hat{K}_0(\tau)|y\rangle &=&\langle x|\exp[-\tau\hat{O}^{(0)}]|y\rangle\\
&=&\exp[-\tau D(x)_{\pm}^{(0)}]\delta (x-y)
\end{eqnarray}
where the last equality is obtained by expanding the exponential and using equation (97).

To summarize, in this subsection, we obtained the analytical expressions of the effective action with fermion-fermion mean-field ansatz, the Euclidean effective action, and the matrix elements of the heat kernel operator. These fundamental quantities should be useful in calculating various correlation functions and physical quantities for this superfluid in curved spacetime. The functional determinant involved in the Euclidean effective action typically gives an infinite quantity. A renormalization procedure is generally needed, typically the zeta-functional renormalization is used.

\section{V. Conclusion}
In this paper, we have derived the basic equations describing superfluidity in curved spacetime. Specifically, we have derived the effective action and the self-consistent equation for the scalar field superfluid case. It was found that the geometry of the scalar superfluid core can not be of spherical shape. A numerical calculation of the gravitational correction to the order parameter was carried out for the sample cases. It was found that, depending on the boundary condition, gravity can either enhance or inhibit the superfluidity. It was also found that even with the weak gravity assumption $\lambda <1$, the effect of gravity on the superfluidity can be as significant as $0.5\lambda -0.8\lambda$. This shows that gravity should not be neglected when treating the formation of superfluids inside neutron stars.

For fermions, two cases were considered: antifermion-fermion and fermion-fermion pairings. The effective actions and the self-consistent equations were derived for both cases. The explicit analytical expression for the first-order gravitational correction to the order parameter was calculated for the antifermion-fermion case. For the fermion-fermion pairing, which is the type of pairing in color-superconducting state and color-flavor locked phase, we further derived the Euclidean effective action and the matrix elements of the heat kernel operator. 

We have also discussed the case of the fermions mediated by the gauge vector bosons. A similar mean-field approach that we did in Section IV can also be done for this case. For quark-antiquark pairing, we can define the non-vanishing order parameter as $j_{\mu}(x)\equiv g\bar{\psi}(x)\gamma_{\mu}(x)T^c_3\psi(x)$. For quark-quark pairing, the order parameter is the $4\times 4$ Dirac-spinor matrix $J_{ab}(x,y)\equiv\psi_{Ca}(x)\bar{\psi}_b(y)$, where $\psi_C$ denotes a charge-conjugated field. The resulting action, which is bilinear in the field operators, can then be integrated.  From here, the self-consistent and the gap equations can also be derived. 

Future research on this topic can now be directed to the calculation of various correlation functions and physical quantities by using the relevant expressions and equations derived in this paper. The functional determinant in the Euclidean effective action will usually give an infinite result and so a renormalization procedure is needed. In curved spacetime quantum field theory, the zeta-functional renormalization is usually more convenient to use.

\section*{Appendix}
In the heat kernel method, we introduce the Hilbert space spanned by the coordinate basis $\{|x\rangle \}$ and the operators $\hat{O}_{\pm}$ which have the same eigenvalues as $\hat{d}\pm \hat{g}$, respectively. These basis and operators obey
\begin{eqnarray}
\langle x|x'\rangle &=&\delta (x-x')\\
\int d^4x |x\rangle\langle x| &=& 1\\
\hat{O}|\psi_n\rangle_{\pm}&=&\lambda_{\pm n}|\psi_n\rangle_{\pm}\\
_{\pm}\langle\psi_n|\psi_m\rangle_{\pm}&=&\delta_{mn}
\end{eqnarray}
where $_{\pm}\langle\psi_n|\equiv (|\psi_n\rangle_{\pm})^{\dagger}\gamma_{\hat{0}}$.

In coordinate basis, equations (104) and (105) can be written as
\begin{eqnarray}
\int d^4x'\langle x|\hat{O}_{\pm}|x'\rangle\psi_n(x')_{\pm}&=&\lambda_{n\pm}\psi_n(x)_{\pm}\\
\int d^4x\bar{\psi}_n(x)_{\pm}\psi_m(x)_{\pm}&=&\delta_{nm}.
\end{eqnarray}

Comparing these with Eq.(79) and the normalization 
\begin{equation}
\int d^4 x\sqrt{g}\bar{\pi}_n(x)_{\pm}\pi_m(x)_{\pm}=\delta_{mn}
\end{equation}
we have 
\begin{eqnarray}
\psi_m(x)_{\pm}&=&g^{1/4}(x)\pi_m(x)_{\pm}\\
\langle x|\hat{O}_{\pm}|x'\rangle &=& g^{1/4}(x)(\hat{d}\pm\hat{g})[g^{-1/4}(x)\delta (x-x')].
\end{eqnarray}

\end{document}